\documentclass{aa}
\usepackage{natbib}
\usepackage[dvips]{graphicx}
\bibpunct{(}{)}{;}{a}{}{,}
\usepackage{rotating}
\usepackage{longtable}
\begin{document}
\title{The distribution of exoplanet masses}
\author{A. Jorissen\inst{1}\fnmsep\thanks{Research Associate, F.N.R.S., 
Belgium}
\and 
M. Mayor\inst{2}
\and 
S. Udry\inst{2}
}
\institute{Institut d'Astronomie et d'Astrophysique, 
Universit\'e Libre de Bruxelles, C.P.~226, Boulevard du Triomphe, B-1050 
Bruxelles, Belgium
\email{ajorisse@astro.ulb.ac.be}
 \and Observatoire de Gen\`eve, CH-1290 Sauverny, Switzerland
\email{Michel.Mayor,Stephane.Udry@obs.unige.ch}
}
\date{Received date / Accepted date} 
\offprints{A. Jorissen}


\abstract{
The present study derives the distribution of secondary masses $M_2$
for the 67 exoplanets and very low-mass brown-dwarf companions of
solar-type stars, known as of April 4, 2001. This distribution is
related to the distribution of $M_2 \sin i$ through an integral
equation of Abel's type. Although a formal solution exists for this
equation, it is known to be ill-behaved, and thus very sensitive to
the statistical noise present in the input $M_2 \sin i$
distribution. To overcome that difficulty, we present two robust,
independent approaches: (i) the formal solution of the integral
equation is numerically computed after performing an optimal smoothing
of the input distribution, (ii) the Lucy-Richardson algorithm is used
to invert the integral equation.  Both approaches give consistent
results.  The resulting statistical distribution of exoplanet true
masses reveals that there is no reason to ascribe the transition
between giant planets and brown dwarfs to the threshold mass for
Deuterium ignition (about 13.6 M$_{\rm J}$).  The $M_2$ distribution
shows instead that most of the objects have $M_2 \leq 10$~M$_{\rm
J}$, but there is a small tail with a few heavier candidates around
15~M$_{\rm J}$.  
 }

\maketitle

\section{Introduction}

\citet{Han-2001:a} suggested that most of the exoplanet candidates
discovered so far have masses well above the lower limit defined by
$\sin i = 1$ (where $i$ is the inclination of the orbital plane on the
sky) and should therefore be considered as brown dwarfs or even stars
rather than planets.  The present paper shows that this claim is not
consistent with the distribution of masses extracted from the observed
$M_2 \sin i$ distribution (where $M_2$ is the mass of the companion)
under the reasonable assumption that the orbits are oriented at random
in space.  Although the distributions of $M_2$ and $M_2 \sin i$ are
related through an integral equation of Abel's kind
\citep{Chandrasekhar-1950, Lucy-74}, its numerical solution is
ill-behaved.  Two different methods are used here to overcome that
difficulty.  In the first method (Sect.~\ref{Sect:Abel}), the formal
solution of Abel's equation is implemented numerically on an input
$M_2 \sin i$ distribution that has been optimally smoothed with an
adaptive kernel procedure \citep{Silverman-86} to remove the
high-frequency fluctuations caused by small-number statistics. The
other method (Sect.~\ref{Sect:LR}) is based on the Lucy-Richardson
inversion algorithm \citep{Richardson-72,Lucy-74}.

The basic reason why the $M_2$ distribution obtained in
Sect.~\ref{Sect:results} differs from that of \citet{Han-2001:a} is
because these authors concluded that most of the systems containing
exoplanet candidates are seen nearly pole-on. This conclusion, based
on the analysis of the Hipparcos {\it Intermediate Astrometric Data}
(IAD), has however been shown to be incorrect
\citep[e.g.][]{Halbwachs-2000:a,Pourbaix-2001:a,Pourbaix-2001:b}, as
summarized in Sect.~\ref{Sect:discussion}.

While this paper was being referred, \citet{Zucker-Mazeh-2001}
  and \citet{Tabachnik-Tremaine-2001} proposed other interesting
  approaches to derive the exoplanet mass distribution.  
  Zucker \& Mazeh derive the binned true mass distribution 
  by using a maximum likelihood approach to retrieve the average
  values of the mass distribution over the selected bins. Their results are in very
  good agreement with ours. 
  Tabachnik \& Tremaine suppose that   
  the period and mass distributions 
  follow power laws, and derive the corresponding power-law 
  indices from a maximum likelihood method.   
  On the contrary, the methods used in the
  present paper (and in Zucker \& Mazeh's) 
  are {\it non-parametric in
  nature,} since they do not require to make any {\it a priori}
  assumptions on the functional form of the mass
  distribution. This is especially important since the comparison of
  the shapes of the mass distributions for exoplanets and low-mass
  stellar companions may provide clues to identify the process by
  which they formed.  By imposing a power-law function like that
  observed for low-mass stellar companions,
  \citet{Tabachnik-Tremaine-2001} somehow implicitly assume that these
  processes must be similar.

\section{The integral equation of Abel's kind relating the
  distributions of $M_2 \sin i$ and $M_2$}
\label{Sect:Abel}

The $M_2 \sin i$ values for low-mass companions of main sequence stars
may be extracted from the spectroscopic mass function and from the
primary mass as derived through e.g., isochrone fitting. Let $\Phi(Y)$
be the observed distribution of $Y \equiv M_2 \sin i$ which is
easily derived from the observed spectroscopic mass functions provided
that $M_2 << M_1$ as it is expected to be the case for the systems
under consideration.  Then, the seeked distribution $\Psi(M_2)$ obeys
the relation
\begin{equation}
\label{Eq:integral}
\Phi(Y) = \int_0^\infty \Psi(M_2)\; \Pi(Y | M_2)\; {\rm d}M_2.
\end{equation} 
The kernel $\Pi(Y | M_2)$ corresponds to the conditional probability
of observing the value $Y$ given $M_2$. Under the assumption that the
orbits are oriented at random in space, the inclination angle $i$
distributes as $\sin i$, and the following expression is obtained for
the kernel:
\begin{equation}
\Pi(Y | M_2) =  \frac{\sin i_0}  {M_2  \cos  i _0}, 
\end{equation}
where $i_0$ satisfies the relations $M_2 \sin i_0 - Y = 0$ and $0 \le
i_0 \le 90$. Eliminating the inclination $i_0$ in the above relation
yields
\begin{equation}
\label{Eq:Pi}
\Pi(Y | M_2) =  \frac{Y}{M_2} \frac{1}{(M_2^2 - Y^2)^{1/2}}\quad\quad\mbox{\rm with}\quad Y \le M_2.
\end{equation}
Eq.(\ref{Eq:integral}) thus rewrites
\begin{equation}
\Phi(Y)   = Y \int_Y^\infty \Psi(M_2) \; \frac{1}{M_2 (M_2^2 - Y^2)^{1/2}}
\; {\rm d}M_2 .
\label{Eq:integral2}
\end{equation}
Eq.(\ref{Eq:integral2}) is the integral equation to be solved for
$\Psi(M_2)$. It can be reduced to Abel's integral equation by the
substitutions \citep{Chandrasekhar-1950}
\begin{equation}
Y^2 = 1/\eta \hspace{6pt}{\rm and} \hspace{6pt} M_2^2 = 1/t.
\end{equation}
With these substitutions, Eq.(\ref{Eq:integral2}) becomes
\begin{equation}
\label{Eq:Abel}
\phi(\eta) = \int_0^{\eta} \frac{\psi(t)}{(\eta - t)^{1/2}} \; {\rm
  d}t,
\end{equation}
where
\begin{equation}
\phi(\eta) \equiv \Phi(\frac{1}{\sqrt{\eta}}) \hspace{6pt} {\rm and}
\hspace{6pt} \psi(t) \equiv \frac{1}{2\;\sqrt{t}}
\Psi(\frac{1}{\sqrt{t}}).
\end{equation} 
It is well known that the solution of Abel's equation
(Eq.~\ref{Eq:Abel}) is given by
\begin{equation}
\label{Eq:Abelsolution}
\psi(t) = \frac{1}{\pi} \int_0^t
\frac{\partial\phi}{\partial \eta}\frac{1}{(t - \eta)^{1/2}} \; {\rm
  d} \eta + \frac{1}{\pi} \frac{\phi(0)}{\sqrt{t}},
\end{equation}
where $\phi(0) = {\rm lim}_{Y \rightarrow \infty} \Phi(Y) = 0$.

While Eq.~(\ref{Eq:Abelsolution}) represents the formal solution of
the problem, it is difficult to implement numerically, since it
requires the differentiation of the observed frequency distribution
$\Phi(Y)$.  Unless the observations are of high precision, it is well
known that this process can lead to misleading results. To overcome
that difficulty, the observed frequency distribution has been smoothed
in an optimal way (see Appendix) before being used in
Eq.~(\ref{Eq:Abelsolution}). The solution $\Psi(t)$ is then computed
numerically using standard differentiation and integration schemes.

\section{The Lucy-Richardson inversion algorithm applied to Abel's
  integral equation}
\label{Sect:LR}

The Lucy-Richardson algorithm provides another robust way to invert
Eq.~(\ref{Eq:integral2}) \citep[see also][]{Cerf-Boffin-94}.  The
method starts from the Bayes theorem on conditional probability in the
form
\begin{equation}
\Psi(M_2) \; \Pi(Y| M_2) = \Phi(Y) \; R(M_2 | Y),
\end{equation}
where $R(M_2 | Y)$ is the reciprocal kernel corresponding to the
integral equation inverse to the one that needs to be solved
(Eq.~\ref{Eq:integral}):
\begin{equation}
\label{Eq:integralinv}
\Psi(M_2) = \int_0^{M_2} \Phi(Y)\;  R( M_2| Y)\; {\rm d}Y.
\end{equation} 

The reciprocal kernel represents the conditional probability that the
binary system has a companion mass $M_2$ when the observed $M_2 \sin
i$ value amounts to $Y$. Thus, one has:
\begin{eqnarray}
R(M_2|Y) &=& \frac{\Psi(M_2) \; \Pi(Y|M_2)}{\Phi(Y)} \\
         &=& 
\frac{\Psi(M_2) \;  \Pi(Y|M_2)}{\int_0^{\infty} \Psi(M_2)\; \Pi(Y|M_2)\; {\rm d}M_2},
\end{eqnarray}
which obviously satisfies the normalization condition $\int_0^{\infty}
R(M_2|Y)\; {\rm d}M_2 = 1$. The problem in solving
Eq.~(\ref{Eq:integralinv}) arises because $R(M_2|Y)$ also depends on
$\Psi(M_2)$, so that an iterative procedure must be used. If
$\Psi_r(M_2)$ is the $r$th estimate of $\Psi(M_2)$, it can be used to
obtain the $(r+1)$th estimate in the following way:
\begin{equation}
\label{Eq:1}
\Psi_{r+1}(M_2) = \int_0^{M_2} \Phi(Y)\; R_r(M_2|Y)\; {\rm d}Y
\end{equation}
with
\begin{equation}
\label{Eq:2}
R_r(M_2|Y) =  \frac{\Psi_r(M_2)\; \Pi(Y|M_2)}{\Phi_r(Y)} 
\end{equation}
and
\begin{equation}
\Phi_r(Y) = \int_0^{\infty} \Psi_r(M_2)\; \Pi(Y|M_2)\; {\rm d}M_2.
\end{equation}
Thus, $\Phi_r(Y)$ represents the corresponding $r$th estimate of the
observed distribution $\Phi(Y)$. Eqs.(\ref{Eq:1}) and (\ref{Eq:2})
together yield the recurrence relation for the $\Psi_r$'s,
\begin{equation}
\label{Eq:recurrence}
\Psi_{r+1}(M_2) = \Psi_r(M_2) \; \int_0^{M_2} \frac{\Phi(Y)}{\Phi_r(Y)}\; \Pi(Y| M_2)\; {\rm d}Y,
\end{equation}
with $\Pi(Y|M_2)$ given by Eq.~(\ref{Eq:Pi}) for the problem under
consideration.  The conditions for convergence of this recurrence
relation are discussed by \citet{Lucy-74} and
\citet{Cerf-Boffin-94}. It needs only be remarked here that (i) the
iterative scheme converges if $\Phi_r(Y)$ tends to $\Phi(Y)$, given
the normalization of the probability $\Pi(Y|M_2) {\rm d}Y$, and (ii)
the full convergence of the method is not necessarily desirable, as
the successive estimates $\Phi_r(Y)$ will tend to match $\Phi(Y)$ on
increasingly smaller scales, but the small-scale structure in
$\Phi(Y)$ is likely to be dominated by the noise in the input
data. This is well illustrated in Fig.~\ref{Fig:2} below.

When the number of data points is small \citep[typically $N <
100$;][]{Cerf-Boffin-94}, it is advantageous to express $\Phi(Y)$ as
\begin{equation}
\Phi(Y) = \frac{1}{N}\sum_{n = 1}^{N} \delta(Y - y_n)
\end{equation}
where the $y_n \; (n=1,\dots N)$ are the $N$ individual measured $M_2
\sin i$ values and $\delta(x)$ is the Dirac `function' such that $x_0
= \int \delta (x - x_0)\; {\rm d}x$.  Substitution in Eq.~(\ref{Eq:1})
then yields
\begin{equation}
\Psi_{r+1}(M_2) = \frac{1}{N}\sum_{n=1}^N R_r(M_2|y_n)
\end{equation}
where $R_r(M_2|y_n)$ is defined as in Eq.~(\ref{Eq:2}). The sample
size should nevertheless be large enough for the functions
$R_r(M_2|y_n)$ to have sufficient overlaps so as to produce a smooth
$\Psi_{r+1}(M_2)$ function.

In the application of the method described in
Sect.~\ref{Sect:results}, the initial mass distribution $\Psi_0(M_2)$
was taken as a uniform distribution, but it has been checked that this
choice has no influence on the final solution.

\section{The frequency distribution $\Psi(M_2)$}
\label{Sect:results}

\begin{figure}
\resizebox{10cm}{8cm}{\includegraphics{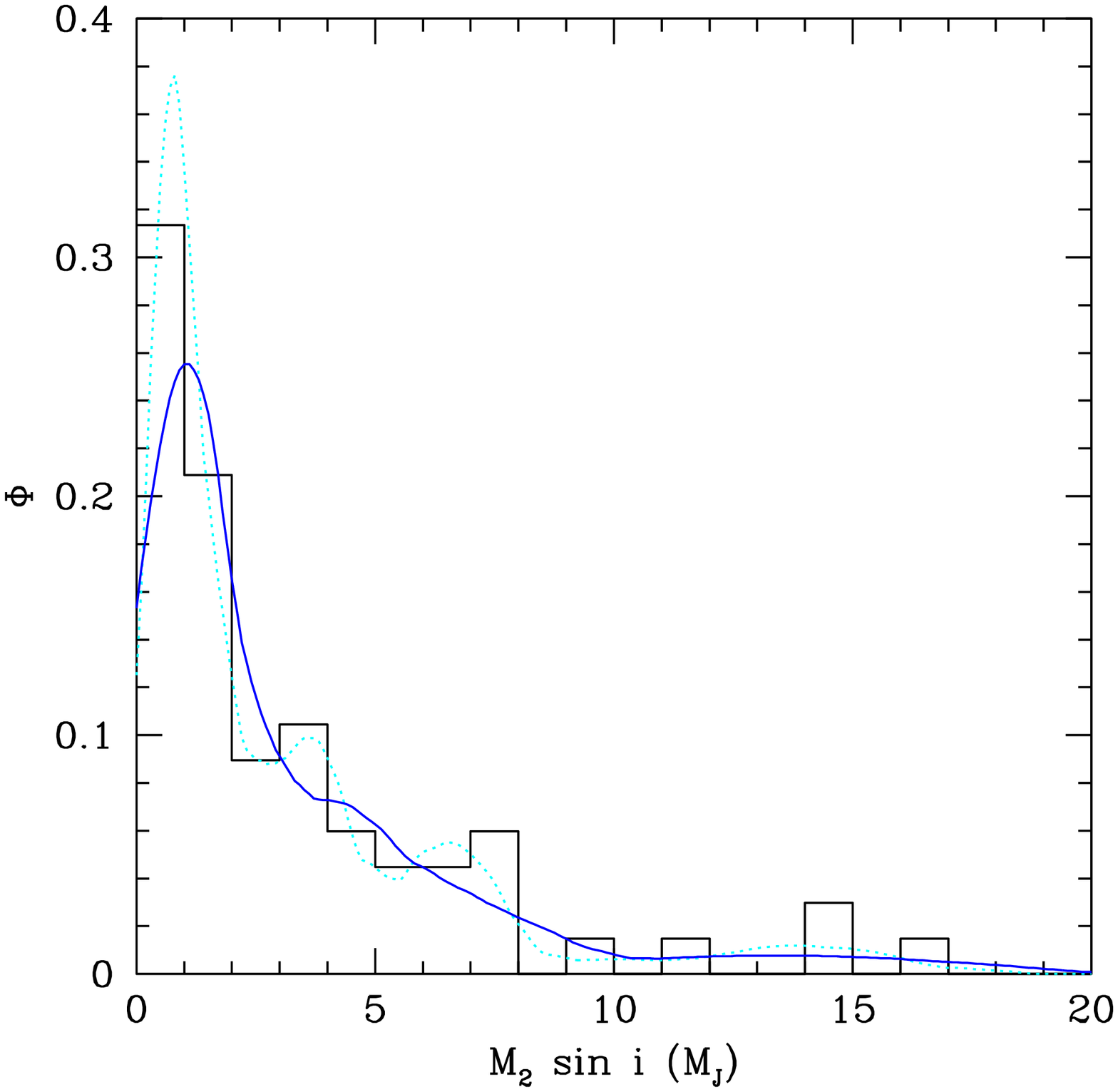}}
\resizebox{10cm}{8cm}{\includegraphics{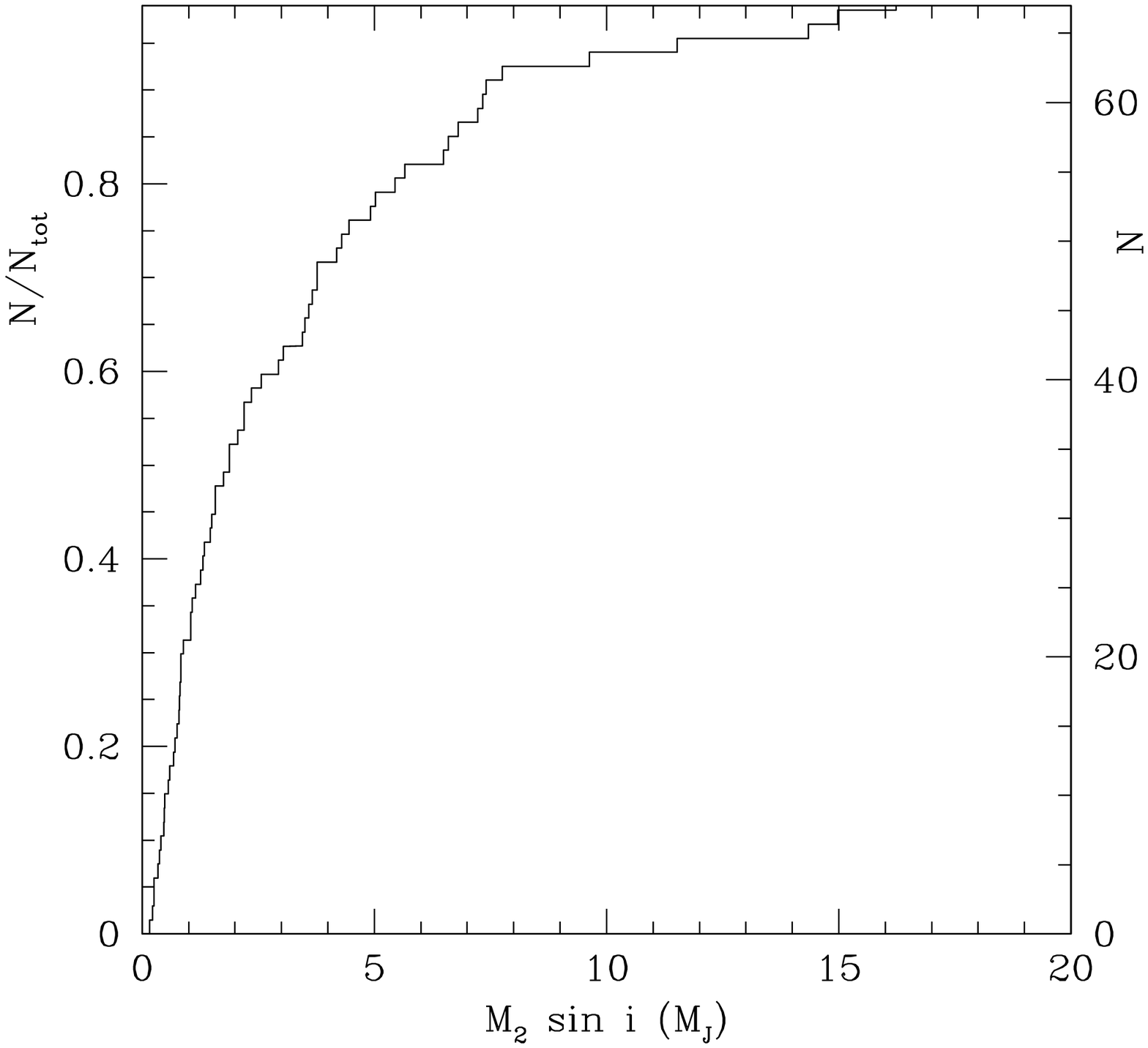}}
\caption[]{
\label{Fig:1}
Bottom panel: The cumulative frequency distribution of the observed
$M_2 \sin i$ values for the 67 exoplanets and very low-mass brown
dwarfs (around 60 stars) known as of April 4, 2001. Top panel: The
corresponding frequency distribution, represented either as an
histogram, or smoothed using an Epanechnikov kernel (see Appendix)
with an optimal smoothing length of $h_{\rm opt} = 1$~M$_{\rm J}$
(dashed curve). The frequency distribution smoothed with $2 h_{\rm
opt}$ is also shown for comparison (solid curve) }
\end{figure}

The cumulative frequency distribution of the $M_2 \sin i$ values
smaller than 17\,M$_{\rm J}$, where M$_{\rm J}$ is the mass of
Jupiter (= M$_\odot$ /1047.35), available in the literature (as of April
4, 2001) is presented in Fig.~\ref{Fig:1}. It appears to be
sufficiently well sampled to attempt the inversion procedure. The
corresponding frequency distributions smoothed with two different
smoothing lengths, locally self-adapting around $h_{\rm
opt}=1$\,M$_{\rm J}$ and 2$h_{\rm opt}$ (see Appendix) are presented
as well for comparison.

The sample includes 60 main-sequence stars hosting 67 companions with
$M_2 \sin i < 17$~M$_{\rm J}$.  Among those, 6 stars are orbited by
more than one companion, namely \object{HD\,74156} and
\object{HD\,82943}\ \citep[$M_2 \sin i = 1.50$,
7.40~M$_{\rm J}$, and $M_2 \sin i =0.84$, 1.57 M$_{\rm J}$,
respectively;][]
{Udry-2001}\footnote{see 
also {\tt
http://www.eso.org/outreach/press-rel/pr\-2001/pr-07-01.html}}, 
\object{HD\,83433} \citep[0.16,
0.38 M$_{\rm J}$;][]{Mayor-2000}, 
\object{HD\,168443} \citep[7.22, 16.2~M$_{\rm
  J}$;][]{Udry-2000:b}, 
\object{$\upsilon$~And}
\citep[0.71, 2.20 and 4.45~M$_{\rm J}$;][]{Butler-99} and 
\object{Gliese 876} 
\citep[0.56, 1.88~M$_{\rm J}$;][]{Marcy-2001}.  
The inversion process is only able to treat these systems under
the hypothesis that the orbits of the different planets in a given
system are {\it not} coplanar, since Eq.~(\ref{Eq:integral2}) to hold
requires random orbital inclinations.  The case of coplanar and non-coplanar 
orbits are discussed separately in the remaining of this section.

\subsection{Non-coplanar orbits}

\begin{figure}
{\resizebox{\hsize}{!}{\includegraphics{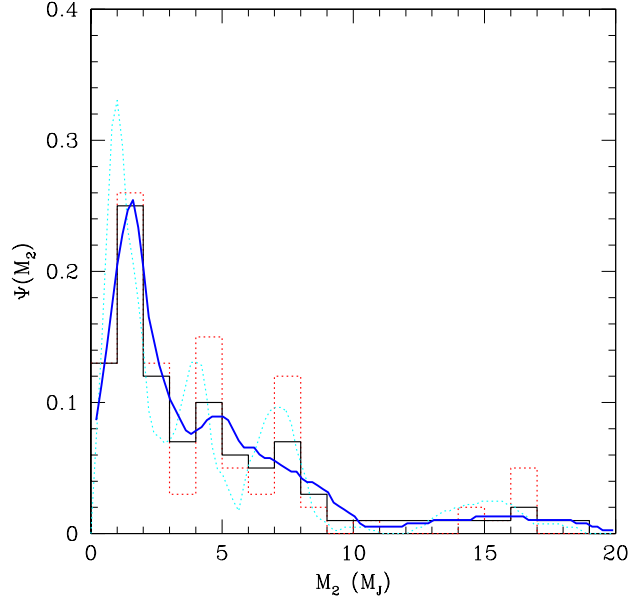}}}
\caption[]{
\label{Fig:2}
Comparison of $\Psi(M_2)$ distributions obtained from the
Lucy-Richardson algorithm (after 2 and 20 iterations, represented by
the solid and dashed histograms, respectively), and from the formal
solution of Abel's integral equation with smoothing lengths $2h_{\rm
opt}$ and $h_{\rm opt}$ applied on $\Phi(Y)$ (represented by solid and
dashed curves, respectively). The supposedly best representation of
the actual $\Psi(M_2)$ distribution is represented by the thick solid
line.  Planetary systems have been included in the inversion process
with the assumption of non-coplanarity of the planetary orbits in a
given system }
\end{figure}

Figure~\ref{Fig:2} compares the solutions $\Psi(M_2)$ obtained from
the Lucy-Richardson algorithm (after 2 and 20 iterations, denoted
$\Psi_2$ and $\Psi_{20}$, respectively) and from the formal solution
of Abel's integral equation with smoothing lengths $h_{\rm opt}$ and 2
$h_{\rm opt}$ on $\Phi(Y)$ (the corresponding solutions are denoted
$\Psi_{h_{\rm opt}}$ and $\Psi_{2h_{\rm opt}}$). The solutions from
the two methods basically agree with each other, although solutions
with different degrees of smoothness may be obtained with each method.
On the one hand, $\Psi_{20}$ and $\Psi_{h_{\rm opt}}$ exhibit
high-frequency fluctuations that may be traced back to the statistical
fluctuations in the input data. This can be seen by noting that the
peaks present in $\Psi_{h_{\rm opt}}$ correspond in fact to the
high-frequency fluctuations already present in $\Phi_{h_{\rm opt}}$
(Fig.~\ref{Fig:1}). These fluctuations should thus not be given
much credit. The same explanation holds true for $\Psi_{20}$,
since it was argued in Sect.~\ref{Sect:LR} that the solutions $\Psi_r$
resulting from a large number of iterations tend to match $\Phi$ at
increasingly small scales (i.e., higher frequencies) where statistical
fluctuations become dominant.  On the other hand, $\Psi_{2}$ and
$\Psi_{2h_{\rm opt}}$ are much smoother, and are probably better
matches to the actual distribution. The local maximum
around $M_2 \sim 1$~M$_{\rm J}$ is very likely however 
an artifact of the strong
detection bias against low-mass companions.

The most striking feature of the $\Psi(M_2)$ distribution displayed in
Fig.~\ref{Fig:2} is its decreasing character, reaching zero for the
first time around $M_2 = 10$~M$_{\rm J}$, and in any case well before
13.6 $M_{\rm J}$. The latter value, corresponding to the minimum
stellar mass for igniting Deuterium, does not in any way mark the
transition between giant planets and brown dwarfs, as sometimes
proposed.  That transition, which is thus likely to occur at smaller
masses, must rely instead on the different mechanisms governing the
formation of planets and brown dwarfs.  Another argument favouring a
giant-planet/brown-dwarf transition mass smaller than 13.6~M$_{\rm J}$
is provided e.g., by the observation of {\it free-floating} (and thus
most likely {\sl stellar}) objects with masses probably smaller than
10~M$_{\rm J}$ in the $\sigma$ Orionis star cluster
\citep{Zapatero-2000}.  The $\Psi(M_2)$ distribution nevertheless
clearly exhibits a tail of objects clustering around $M_2 \sim
15$~M$_{\rm J}$, due to \object{HD\,114762} ($M_2 \sin i =
11.5$~M$_{\rm J}$), \object{HD\,162020} (14.3~M$_{\rm J}$),
\object{HD\,202206} (15.0~M$_{\rm J}$) and \object{HD\,168443\,c}
(16.2~M$_{\rm J}$).  It would be interesting to investigate whether
these systems differ from those with smaller masses in some
identifiable way (periods, eccentricities, metallicities,...), so as
to assess whether or not they form a distinct class (Udry et al., in
preparation).
 
\begin{figure}
{\resizebox{\hsize}{!}{\includegraphics{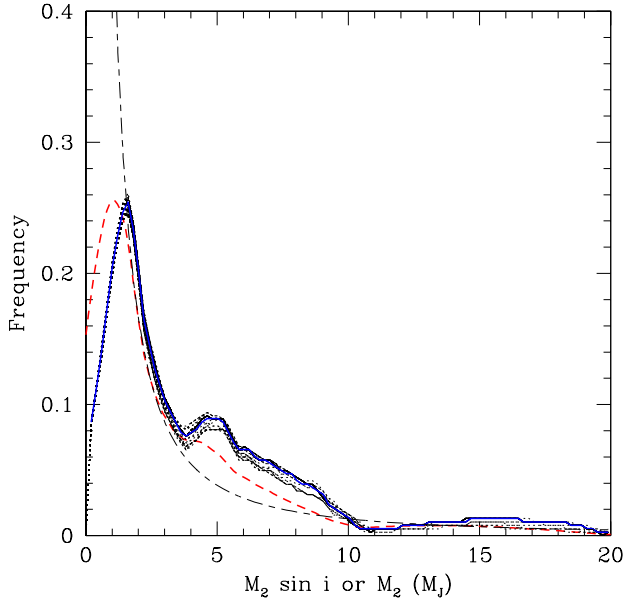}}}
\caption[]{
\label{Fig:jackknife}
Comparison of the input $\Phi_{2h_{\rm opt}}$ distribution (thick dashed
line) and the 
67 $\Psi_{2h_{\rm opt}}$  distributions (thin dotted lines) 
resulting from the application of 
the jackknife method (see text), which illustrates the uncertainty on
the solution $\Psi_{2h_{\rm opt}}$ (thick solid line). To guide the
eye, a power-law of index $-1.6$ has been plotted as well (dot-dashed line)
}
\end{figure}

The jackknife method \citep[e.g.,][]{Lupton-93} has been used to
  estimate the uncertainty on the $\Psi_{2h_{\rm opt}}$ solution. In a
  first step, 67 input $\Phi_{2h_{\rm opt}}$ distributions are
  computed, corresponding to all 67 possible sets with one data point
  removed from the original set. Eq.~\ref{Eq:Abelsolution} is then
  applied to these 67 different input distributions. The resulting
  distributions are displayed in Fig.~\ref{Fig:jackknife}, which shows
  that the threshold observed at 10~M$_{\rm J}$ is a robust result not
  affected by the uncertainty on the solution.

\subsection{Coplanar orbits in multi-planets systems}

\begin{figure}
\resizebox{\hsize}{!}{\includegraphics{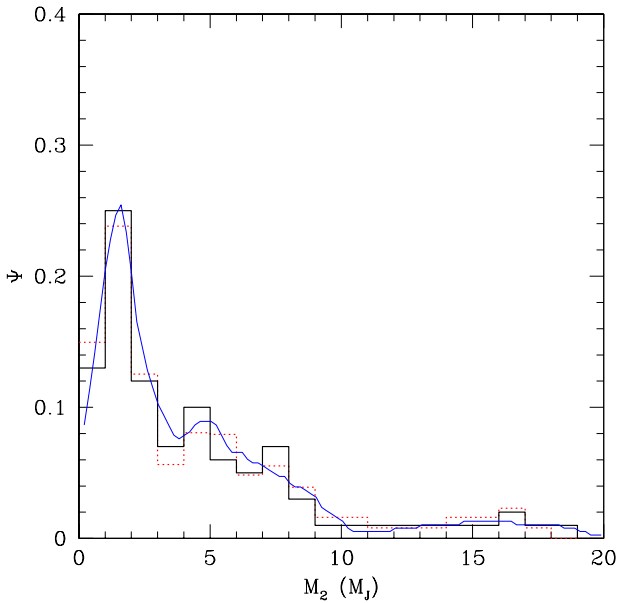}}
\caption[]{
\label{Fig:coplanar}
Evaluation of the impact of the coplanarity hypothesis on the
resulting $\Psi(M_2)$ distribution. The dashed histogram corresponds
to the mass distribution obtained assuming coplanar orbits in
planetary systems (see text), as compared to the $\Psi_2$ (solid
histogram) and $\Psi_{2h_{\rm opt}}$ solutions (see
Fig.~\protect\ref{Fig:2}) }
\end{figure}

All the results discussed so far are obtained under the assumption
that orbits of planets belonging to a planetary system are {\it not}
coplanar.  To evaluate the impact of this hypothesis, the following
procedure has been applied.  In a first step, the Lucy-Richardson
algorithm is applied on the data set excluding the 13 planets
belonging to planetary systems. That mass distribution obtained after
2 iterations is then completed by mass estimates for the remaining 13
planets. For each of the 6 different systems, an inclination $i$ is
drawn from a $\sin i$ distribution.  This is done through the
expression $i = {\rm acos} \; x$, where $x$ is a random number with
uniform deviate. The same value of $i$ is then applied to all planets
in a given system to extract $M_2$ from the observed $M_2 \sin i$
value.  The distributions of exoplanet masses obtained with and
without the hypothesis of coplanarity are compared in
Fig.~\ref{Fig:coplanar}, and it is seen that planetary systems are not
yet numerous enough for the coplanarity hypothesis to alter
significantly the resulting $\Psi(M_2)$ distribution.

In any case, the main result of the present paper is that {\it
the statistical properties of the observed $M_2 \sin i$ distribution
coupled with the hypothesis of randomly oriented orbital planes
confine the vast majority of planetary companion masses below about 10 $M_{\rm J}$}.
\citet{Zucker-Mazeh-2001} reach the same conclusion.

It should be remarked that the above conclusion cannot be due to
detection biases, since the high-mass tail of the $M_2$ distribution 
is not affected by the 
difficulty of finding low-amplitude, long-period orbits.

\section{Discussion}
\label{Sect:discussion}

Although the assumption of random orbital inclinations seems
reasonable, it is at variance with the conclusion of
\citet{Han-2001:a} that most of the systems containing exoplanet
candidates are seen nearly pole-on.  These authors reached this
conclusion by trying to extract the astrometric orbit, hence the
orbital inclination, from the Hipparcos IAD. \citet{Halbwachs-2000:a}
had already cautioned that this approach is doomed to fail for systems
with apparent separations on the sky that are below the Hipparcos
sensitivity (i.e. $\la 1$~mas). In those cases, the solution retrieved
from the fit of the IAD residuals is spurious, since the true angular
semi-major axis $a$ is simply too small to be seen by Hipparcos. Since
\citet{Halbwachs-2000:a} have shown that $a$ actually follows a
Rayleigh probability distribution, the fit of the IAD residuals will
yield a solution larger than the true value, in fact of the order of
the residuals. But since $a \sin i$ is constrained by the
spectroscopic orbital elements, the too-large astrometric $a$ value
will force $i$ to be close to $0$ to match the spectroscopic value of
the product $a \sin i$, as convincingly shown by
\citet{Pourbaix-2001:a}.  Hence, this approach gives the impression
that all orbits are seen nearly face-on. As an illustrative example,
\citet{Pourbaix-2001:b} have shown that such an approach leads to a
stellar mass for the companion of \object{HD\,209458} that, on another hand,
has been proven to be a 0.69\,M$_{\rm J}$ planet by the photometric
observation of the planet transit in front of the star
\citep{Charbonneau-2000}.

The \citet{Han-2001:a} result is moreover statistically very
unlikely if the orbital planes are oriented at random in space
\citep{Pourbaix-2001:b}. \citet{Han-2001:a} have tried to justify this
unlikely statistical occurrence by invoking biases against
high-amplitude orbits in the selection
process of the radial-velocity-monitoring samples.  To the contrary, 
the planet-search surveys were specifically devised to avoid such biases, as 
they aim at finding not only giant planets but also brown dwarfs 
so as to constrain the substellar secondary mass function
of solar-type stars.
Furthermore, \citet{Han-2001:a} argument
is totally invalid in the case of volume-limited, statistically well-defined
samples like that of the CORALIE planet-search programme in the southern
hemisphere \citep{Udry-2000:a}. This sample has been specifically 
designed to detect companions
of solar-type stars all the way from $q=M_2/M_1 = 1$ down to $q\leq 0.001$.

\section*{Appendix: Non-parametric treatment of the data}

To decrease the noise and allow a tractable use of the information
present in small data samples, heavy smoothing techniques are often
required.  A common practice consists in converting a set of discrete
positions into binned ``counts''.  Binning is a crude sort of
smoothing and many studies in statistical analysis have shown that,
unless the smoothing is done in an optimum way, some, or even most, of the
information content of the data could be lost. This is especially true
when a large amount of smoothing is necessary, which then changes the
``shape'' of the resulting function. In statistical terms, the
smoothing process not only decreases the noise (i.e., the 
function's variance), but at the same time
introduces a bias in the estimate of the function. 

The variance-bias trade off is unavoidable but, for a given level of
variance, the bias increase can be
minimized.  The correct manner of
achieving that task is provided by the so-called {\sl non-parametric
density estimate} methods for the determination of the ``frequency''
function of a given parameter or by the {\sl non-parametric
regression} methods for the determination of a smooth function $g$
inferred from direct measurements of $g$ itself. Moreover, {\sl
adaptive} non-parametric methods are designed to filter the data in
some local, objective way minimizing the bias, in order to get the
smooth desired function without constraining its form in any way.
The theory and algorithms related to those methods, originally built 
to handle ill-conditioned problems (either under-determined or extremely
sensitive to errors or incompleteness in the data), are widely
discussed in the specialized literature and summarized in easy-to-read
textbooks \citep[e.g.,][]{Silverman-86,Hardle-90,Scott-92}.

The simplest of the available algorithms is provided by the kernel
estimator leading to the following form of the normalized
``frequency'' function
\begin{equation}
\label{Eq:frequency}
\hat f_K(x) = {1\over N h}\sum_{n=1}^N K\left({x-X_n\over h}\right) 
\end{equation}
with the normalization
\begin{equation}
\int_{-\infty}^{\infty} K(u)\;{\rm d}u = 1,
\end{equation}
where $X_n$ ($n=1,...N$) are the $n$ available 
data points.  Each data point is thus simply replaced by a
`bump'.  The {\sl kernel function} $K$ determines the shape of the bumps
while the {\sl bandwidth} $h$ (also called {\sl smoothing parameter})
determines their width\footnote{In the regression problem, the
Nadaraya-Watson estimator \citep{Nadaraya-64, Watson-64} is commonly
used:
\begin{displaymath}
g(x) = \frac{\displaystyle \sum_{n=1}^N{g(X_n) K(\frac{x-X_n}{h})}}
{\displaystyle \sum_{n=1}^N{K(\frac{x-X_n}{h})}} 
\end{displaymath}
}.

In the {\sl adaptive kernel} version, a local bandwidth
$h_n=h(X_n,f)$ is defined and used in Eq.~\ref{Eq:frequency}. In order to
follow the ``true'' underlying function in the best possible way, the
amount of smoothing should be small when $f$ is large whereas more
smoothing is needed where $f$ takes lower values.  A convenient method 
to do so consists in deriving first a pilot estimate $\tilde f$ 
of $f$, e.g. by
using an histogram or a kernel with fixed bandwidth $h_{opt}$, and
then by defining the local bandwidths
\begin{equation}
 h_n=h(X_n)=h_{opt}[\tilde f(X_n)/s]^{-\alpha},
\end{equation}
where
\begin{equation}
\log{s}=\frac{1}{N}\sum_{n=1}^N{\log{\tilde f(X_n)}}.
\end{equation}
It may be shown \citep{Abramson-82} that $\alpha=1/2$ leads to a bias of
smaller order than for the fixed $h$ estimate.

The optimum kernel $K$ may be taken as the one minimizing the integrated
mean square error beween $f$ and $\hat f$ (MISE), where
\begin{equation}
MISE(\hat{f}) = E\int\!\left[ \hat{f}(x) - f(x) \right]^2 {\rm d}x
\end{equation}
is usually taken as an indicator of efficiency. In the above
expression, $E$ denotes the statistical expectation. This optimum kernel 
is the so-called
Epanechnikov ($K_e$) or quadratic kernel:
\begin{equation}
K_e(u) = {3\over 4}(1-u^2),\ \ \  |u|<1.
\end{equation}
Other choices of $K$ differ only slightly in their asymptotic
efficiencies and could be more adapted to particular purposes as e.g.,
for bi-dimensional data.

The pilot smoothing length ($h_{opt}$) is the only subjective
parameter required by the method. It relates to the quality of the
sampling of the variable under consideration. There are several ways for
automatically estimating an optimum value of $h_{opt}$ \citep[see
e.g.][ for an extensive review]{Silverman-86}. A simple approach based
on the data variance gives in our case $h_{opt}\simeq 1.0$\,M$_{\rm
J}$. As the derivative of the frequency function rather than the
function itself is actually used in
Eq.~(\ref{Eq:Abel}), a larger pilot smoothing length ($2h_{opt}$) was
also considered in order to remove spurious small-statistics
fluctuations of the density estimate.

\bibliographystyle{apj} 

\bibliography{/home/bibtex/ajorisse_articles}

\end{document}